\documentclass[12pt]{JHEP}

\usepackage{epsfig}
\usepackage{amsmath}
\usepackage{amssymb,amsfonts}

\advance\hoffset by -.35in
\preprint{hep-th/0112145\\DAMTP-2001-104}

\title{\Large\bf Fractional D3-branes in diverse backgrounds}

\author{P. Bain\\

\vskip 24pt

Department of Applied Mathematics and Theoretical Physics,\\
University of Cambridge,\\
Wilberforce Road, CB3 0WA Cambridge, U.K.\\}

\abstract{
In the first part of this article, we find fractional D3-branes
supergravity solutions on orientifolded $\mathbb{C}^2/\mathbb{Z}_2$
orbifolds of 
type IIB string theory. The one-loop corrected gauge couplings for the
symplectic or orthogonal groups living on the D-branes are reproduced
on the world-volume of probes. In the second part of the paper, we
construct a D3-brane solution on the two-centers Taub-NUT manifold
which interpolates between 
fractional D3-branes in the ALE space limit and a T-dual smeared type IIA
configuration. Then, we lift this 
configuration to M-theory and comment on the connections with wrapped
M5-branes solutions. 
}


\begin{document}

\section{Introduction}

During the last couple of years, the study of gauge/gravity duality
has partly shifted from the original AdS/CFT correspondence \cite{mal}
and its 
less supersymmetric cousins to more realistic, non-conformal
cases. Different approaches has been proposed and explored, with 
configurations preserving ${\cal N}=2$, ${\cal N}=1$ or no
supersymmetries~: 
mass deformations of a conformal field theory \cite{gppz,ps,pw}, NS5-branes
or 
M5-branes wrapped on a Riemann surface \cite{mn,gkmw,bcz}, fractional
D3-branes 
on orbifold spaces or conifolds
\cite{kn,kt,ks,bgz,bdflmp,pol,aha,prz,bgl,grp,bcffst,bdflm} to cite several
of those constructions. Typically, 
the ``naive'' supergravity solution displays naked
singularities. However, the singularities are usually the 
manifestation of interesting and non-trivial phenomena in the infrared of
the dual gauge theory and their resolutions in supergravity should
take into account such effects. A typical example is the ${\cal N}=1$
solution of \cite{ks}, where the chiral symmetry breaking of the field
theory in the IR removes the singularity by deforming the conifold. 

This example uses fractional D3-branes which first appeared the
context of string theory defined on orbifold spaces \cite{dm}. On a
conifold, one does not have anymore a microscopic description of the  
fractional D-branes and must rely on the pure supergravity solutions. 
One the other hand, for fractional D3-branes in type IIB theory on the
orbifold space $\mathbb{C}^2/\mathbb{Z}_N$, one can use the conformal
field theory technology. Indeed, these D-branes have an exact
description in string theory as boundary states \cite{dg}. Despite being less
interesting for real world applications than their ${\cal N}=1$
counterparts, these D3-branes configurations, whose world-volume gauge
theory possesses an ${\cal N}=2$ supersymmetry, could be seen as tools
to study and to test the gauge/gravity correspondence away from its
conformal 
limit, like the original, maximally (super-)symmetric D3-branes were
useful for the conformal case. Indeed, the larger amount of preserved
supersymmetries gives a better control of the theories on both sides; in
the gauge theory, we still have non-renormalization theorems, telling
us, for instance, that the gauge couplings only receive perturbative
corrections at one-loop. Moreover, its non-perturbative corrections
could be, in principle, obtained using the Seiberg-Witten curve
of the gauge theory \cite{sw}. On the gravity side, one could also expect that
the $\alpha^\prime$ corrections are better controlled, as it is well
known for solutions of heterotic string theories. Moreover, these
${\cal N}=2$ 
configurations display on their own interesting phenomena like the
enhan\c{c}on mechanism \cite{jpp} which was argued to solve the singular
behaviour of their metrics; this characteristic is a common feature of
several kinds of solutions preserving ${\cal N}=2$ supersymmetries
presented in the literature, namely, mass deformation of an ${\cal
N}=4$ gauge theory \cite{bpp,ejp}, NS5-branes wrapped on a Riemann surface
\cite{bcz} or fractional D3-branes \cite{bdflmp,prz,bgl}. Actually,
the enhan\c{c}on mechanism 
is the gravity manifestation of the existence of Landau poles in the
gauge theory, when the one-loop correction cancels the tree-level
contribution. However, we know that the complete metric on the moduli
space of a ${\cal N}=2$ gauge theory is not singular when 
Yang-Mills instantons corrections are included. It is still a challenge to
describe precisely these effects from the dual point-of-view. 

Among ${\cal N}=2$ non-conformal configurations, the  fractional
D3-branes one 
seems to be the simplest. For instance, it is obvious to describe  
an arbitrary point on the Coulomb branch of the gauge theory since the
{\it vev} of the adjoint scalars which parameterize 
the moduli space are proportional to the positions of the fractional
D3-branes and are chosen arbitrarily to construct  the
solution. One the other hand, in mass deformed or wrapped NS5-branes
configurations, the precise identification of the point on the Coulomb
branch requires some work and, so far, it is not clear that any point
can be obtained by deforming the known solutions\footnote{An example
of such deformation have been given in \cite{bcz}, where the {\it vev}
are linearly distributed.}.  
Moreover, we can also see a fractional D-brane configuration as a
``mass deformed'' conformal theory~: one can always start with a stack of
bulk 
D-branes and give masses to bifundamental hypermultiplets by pulling
away some fractional D-branes. At large distance, the theory will be
conformal but, in the IR of the gauge theory, the D-branes sources
break the conformal symmetry. 

The aim of this article, which is made of two independent parts, is
two-folds. In the first part, we will generalize fractional D3-branes
solutions to models with orientifolds. The presence of O-planes leads
to symplectic or orthogonal gauge groups on the world-volume of the
D-branes. After a short review of some basic facts on orientifold
projections and existence of fractional D-branes, we describe the
supergravity solutions for  two possible cases. Such solutions are
similar to their parents type IIB and display common features,
namely singularities and enhan\c{c}on; as expected, we show that the
one-loop corrected gauge couplings for orthogonal and symplectic gauge
groups are reproduced on the world-volume of fractional D3-branes
probing these backgrounds. This result should be interpreted as a
manifestation of the closed/open string duality. We also speculate on
the relation between periodic monopoles and a refined 
solution close to  the enhan\c{c}on radius. 

The second part of this paper is a first step toward a 
comparison of the alternative approaches to describe pure
$SU(N)$ ${\cal N}=2$ gauge theory using supergravity. One way to
study the possible connection between very different branes
configurations is to uplift them to eleven dimensions where they
should be described as M5-branes wrapped on manifolds in such a way
that, in the IR, the world-volume theory flows to the
four-dimensional gauge theory. Moreover, the M-theory point-of-view
could be useful to understand the approximations made and the domains
of validity of the different solutions. In particular,
one can think about non-perturbative effects that are inaccessible in
the type II description. Indeed, we know that such an eleven
dimensional approach has been fruitful in the past to recover 
non-perturbative - D-instantons - corrections to string effective
actions whose form was conjectured using symmetry - self-duality -
arguments \cite{gg,ggv,bbg}. In this 
article, we study the case of the fractional D3-branes, first by
constructing a solution on a two-centers Taub-NUT space which
interpolates between the type IIB fractional D3-branes solution in the
ALE limit and a partially smeared type IIA configuration in another
limit. Then, we lift the solution to M-theory, where it describes an
M5-brane  
wrapped on a Riemann surface. Finally, we will comment on the
relation with M5-branes solutions proposed by \cite{bfms} to be dual 
to the pure $SU(N)$ ${\cal N}=2$ gauge theory but which has received
much less attention so far  than fractional D3-branes and wrapped
NS5-branes. 

{\it Note added in proof~:} while this article was reaching its final
stage, appeared the preprint 
\cite{dveil}, which develops ideas similar to the ones considered in
the second part by comparing fractional D-branes to wrapped branes
solutions in the context of three dimensional gauge theories.

\section{Fractional D-brane solutions with orthogonal and symplectic gauge
groups} 

\subsection{Orientifold projections and fractional D3-branes}

We start with type IIB string theory on the non-compact orbifold space
$\mathbb{C}^2/\mathbb{Z}_2$; the directions parallel to the orbifold
will be called $x^6, x^7, x^8, x^9$. 
The twisted fields of this theory, namely the NS-NS scalar $b$, the three
blow-up modes, $\xi^\pm, \xi^3$, the R-R scalar $c$ and the R-R 2-form
$A^{(2)}$ fill a six dimensional ${\cal N}_6=(2,0)$ tensor
multiplet. In order to discuss the orientifold projection, it is
convenient to use the ${\cal N}_6=(1,0)$
supersymmetry language where this tensor multiplet is split into a 
tensor multiplet and a hypermultiplet. The fields
$b$ and $A^{(2)}$ make up the first while the four other scalars fill
the hypermultiplet. 

``Standard'' orientifold projection \cite{bs,gp} keeps the
hypermultiplet and 
projected out the tensor multiplet. Therefore, the fractional D-branes
which appear is this model always come in pairs. One can see this
easily using a T-dual type IIA description. T-dualizing the orbifold
space $\mathbb{C}^2/\mathbb{Z}_2$ along its $U(1)$ direction maps the
$A_1$ singularity 
to a pair of NS5-branes separated along the T-dualized direction,
$x^6$ \cite{ov,ghm}. The orientifold 9-plane splits into two
orientifolds 8-planes located at antipodal points on the circle
parameterized by $x^6$. For the standard choice, each of the NS5-branes
coincides 
with one $O8$-plane. 
Now, we remind the reader that, under T-duality, the twisted fields
are supposed to be mapped to differences between the fields which live
on each NS5-branes and which also fill ${\cal N}_6=(2,0)$ tensor
multiplets \cite{kls}. For instance, the $b$ field is related to the
distance 
between the two NS5-branes along $x^6$ while the blow-up modes are
mapped to the relative positions of these branes along $x^7, x^8$ and
$x^9$. Therefore, for symmetry reasons, we see that, for such orientifold
choice, the field $b$ and $A^{(2)}$ are projected out. 

One the other hand, the above T-dual picture suggests another choice
for the orientifold projection \cite{dp,bz,bi} which will give
fractional D-branes 
similar to their type IIB counterparts \cite{ep,qs}. Indeed, it is
possible to move 
the NS5-branes off the $O8$-planes in the direction $x^6$ and the
symmetry of the problem imposes that the NS5-branes can not be moved
anymore independently in the $x^7, x^8$ or $x^9$
directions. T-dualizing back to type IIB string theory, this means that
the orientifold operation now projects out the hypermultiplet of the
twisted  sector. The orbifold space lacks its blow-up modes but has a
twisted NS-NS $b$ field and its 2-form companion. In the following, we
will always consider this second possibility and construct the
fractional D-branes solutions on such backgrounds. 

To get fractional D3-branes, we perform two T-dualities along two
directions  
transverse to the orbifold; the R-R 2-form becomes a twisted 4-form
under which the fractional D3-branes will be charged and which can
also be Hodge-dualized to a scalar, called $c$ below. The theory will
contain  $O7$-planes, and 
depending on the orientifold projection, D7-branes to cancel the 
tadpole of the R-R 8-form. More precisely, we will consider in this
paper two different orientifold projections which give rise to
symplectic or/and orthogonal groups on the fractional D3-branes. 

The first one has an orientifold projection defined by
$P_1 \equiv (1+\alpha)(1+\Omega^\prime)/4$ where $\alpha$ is the
generator of the  $\mathbb{Z}_2$ orbifold group acting on the
coordinates as $x^6, x^7, x^8, x^9 \rightarrow - (x^6, x^7, x^8, x^9)$
and $\Omega^\prime = \Omega R_{45}(-1)^{F_L}$ where $\Omega$ is the
standard orientifold operator, $R_{45}$ is the
reflection in $x^4, x^5$ directions and $(-1)^{F_L}$ acts as -1 on the
left-moving Ramond sector. The presence of these last two operators
is due to the T-duality transformations.
This model contains four $O7^-$-planes whose total negative
charge under $C^{(8)}$ is neutralized by adding 32 D7-branes. In this
background, we 
can have fractional D3-branes and the gauge theory which lives on the
world-volume of $N$ of those has a gauge group $Sp(2N)$, with a 
hypermultiplet in the fundamental representation, coming from
strings stretched between the D7-branes and these D3-branes. More
precisely, as in type IIB string theory, we have two different
kinds of fractional D3-branes, whose world-volume theories are obtained
by a projection of the gauge theories living on their type IIB
parents. For the orientifold projection $P_1$ both types of fractional
D-branes have the same massless field content. Therefore, for $2N_1$
fractional D3-branes of the first kind and $2N_2$ of the second kind
(we are counting also their images under $\Omega^\prime$),
the gauge theory will have a group $Sp(2N_1) \times Sp(2N_2)$ with one
hypermultiplet in the bifundamental, two in the $\square_1$ and two
in the $\square_2$.   
The minimally covariantized action of the linear couplings of these
D3-branes to the background fields can be extracted from their boundary
state description~:
\begin{eqnarray}
{\cal S}_{1/2} &=& -  \frac{T_3}{\sqrt{2}\kappa_{\rm orb}} 
\Bigl(\int d^4\sigma\, e^{-\phi}\,
\sqrt{-{\rm det}(g+F)}
\left(1\pm\frac{\tilde{b}}{2\pi^2\alpha^\prime}\right) \nonumber \\
&&~~~~~~~~~~~~~~~~~~~~~~~~~~~~~~~~~~~- \int C^{(4)}
\left(1\pm\frac{\tilde{b}}{2\pi^2\alpha^\prime}\right) 
\pm \frac{A^{(4)}}{2\pi^2\alpha^\prime}\Bigr) 
%
\label{d3action}
\end{eqnarray}
where $g$ is the induced metric on the world-volume of the D-brane~: 
$g_{\alpha\beta} \equiv G_{\mu\nu} \partial_\alpha X^\mu \partial_\beta
X^\nu$. 

The second orientifold projection we will consider does not contain any
D7-branes. The operator is defined by $P_2 \equiv
(1+\alpha)(1+\beta\Omega^\prime)$ where $\beta^2 = \alpha$ and does
not induce any R-R tadpole. In the T-dual picture, the corresponding
orientifold plane splits into two $O6$-planes, one negatively charged and
the other oppositely charged under the R-R 7-form. Again, in this
background, we have two kinds of fractional D3-branes; however, the
orientifold projection now gives rise to a $Sp(2N)$ gauge theory on
the first type and to a $SO(2N)$ gauge 
group on the other. The T-dual description gives again a simple
geometrical 
interpretation of this result~: the first case corresponds to 
D4-branes intersecting the $O6^-$-plane as for $P_1$, while
the second comes from a D4-brane intersecting the positively charged
$O$-plane. The most general configuration will have $N_1$ fractional
D3-branes of the first kind and $N_2$ of the second type (plus their
images under $\Omega^\prime$), leading to an
$Sp(2N_1)\times SO(2N_2)$ gauge theory with one hypermultiplet in the
bifundamental. The superconformal theories constructed using D3-branes
in these models have been studied in \cite{pu,gk}. In the context of
the AdS/CFT correspondence, supergravity solutions for these conformal
configurations have been given in \cite{elns}.

\subsection{Supergravity solutions and one-loop metric on the moduli
space} 

\subsubsection{The field equations}

Since we know their type IIB counterparts, the supergravity
descriptions of 
these fractional D3-branes are easy to find. Indeed, we have just seen
that they couple to
the same supergravity fields. The fields projected out by the
orientifold operators, namely, the untwisted NS-NS 2-form, the
untwisted R-R 2-form, the twisted R-R 2-form and the
blow-up modes were already inactive in the type IIB case. Therefore,
the homogeneous field equations will be identical.
These equations for the scalar fields can be read in
\cite{grp,bcffst}
\begin{eqnarray}
&&d^\star d\tau = 0, \label{taueq} \\
&&d^{\star_6}d \gamma + i F_{(5)} \wedge (dc + \tau db) - b\, d^{\star_6}d
\tau =0 \label{gammaeq}
\end{eqnarray}
where we have introduced the complex combinations
\begin{eqnarray}
\tau = C^{(0)} + i e^{-\phi} \qquad {\rm and } \qquad 
\gamma = c + \tau b \nonumber
\end{eqnarray}
and defined the self-dual 5-form 
\begin{equation}
F^{(5)} = dC^{(4)} + \ ^\star dC^{(4)}. \nonumber
\end{equation}
Requiring that a fraction of the supersymmetries is preserved by the
solution implies that the scalar fields  $\tau$ and $\gamma$ must be
holomorphic functions of $z$. This constraint has been used to 
derive the equations (\ref{taueq}) and (\ref{gammaeq}). The two-dimensional 
homogeneous Laplace equation (\ref{taueq}) is 
automatically satisfied and the precise form of $\tau$ must come from
other arguments, one being that it has to reflect the presence of
sources for the untwisted R-R scalar. An other requirement is that its
imaginary part, $e^{-\phi}$, must stay positive. 
Finally, one can also use its modular properties under the group
$SL(2, \mathbb{Z})$.

Using the standard metric and self-dual 5-form ansatz for D3-branes
on top of D7-branes/O7-plane \cite{afm},
\begin{eqnarray}
ds^2 &=& H^{-1/2}\eta_{\alpha\beta} dx^\alpha dx^\beta + H^{1/2}(e^{\psi} dz
d\bar{z} + \delta_{ij} dx^i dx^j), \nonumber \\
F^{(5)} &=& d(H^{-1} dx^0\wedge \ldots \wedge dx^3) + \ ^\star
d(H^{-1} dx^0\wedge \ldots \wedge dx^3),
\end{eqnarray}
in the Einstein and 5-form equations leads\footnote{In
this article, we will only write the homogeneous 
equations. The presence of the D-brane sources for the fields will be
imposed at the level of the boundary conditions.}
\begin{equation}
 dx^4\wedge dx^5\left(e^{\psi}\delta^{ij} \partial_i \partial_j +
\partial_z\partial_{\bar{z}} \right) H - \delta(x^6)
\ldots \delta(x^9) db \wedge dc 
 = 0
\label{heq}
\end{equation}
for the harmonic function $H(z,\bar{z},x^i)$. Finally, $\psi$ is a
function of 
$z$ and $\bar{z}$ which obeys the equation
\begin{equation}
\partial_z\partial_{\bar{z}} \left(\psi - \log \tau_2\right)=0.
\end{equation}
Compared to the fractional D3-branes on top of D7-branes studied in
\cite{grp,bcffst}, the 
difference lies in the boundary conditions (or, equivalently, source
terms in the right hand side of the above field equations) and in
the geometrical symmetries imposed by the orientifold projections.

\subsubsection{$Sp(2N_1)\times Sp(2N_2)$ gauge theory}

Besides the usual orbifold projection $(1+\alpha)$, the orientifold
projection $P_1$ 
has a geometrical action on the coordinates $z=x^4+i x^5 \rightarrow
-z$, due to the presence of the reflection operator $R_{45}$ in
$\Omega^\prime$. One of the four orientifold 7-plane is located at the
fixed point 
$z=0$ of this operator, the three other being rejected at infinity in
the decompactified limit of the theory. In this study, we will forget
them and consider only the O7-plane at the origin and its eight
D7-branes companions. If one only imposes the global cancellation of
the untwisted R-R tadpole, these D7-branes can be located away
from the 
origin of the $z$-plane. However, the $R_{45}$ symmetry imposes a
constraint on their positions. Moreover, these D7-branes carry a
charge under the twisted R-R field which, in absolute value, is one
quarter of the charge of a fractional D3-brane. Before doing the
T-dualities along $x^4$ and $x^5$, twisted 
R-R tadpole cancellation was a requirement for the consistency of the
theory and imposed that one half of the D9-branes carries a positive
charge and the other half a negative charge. 

In the decompactified theory obtained after T-duality, this
choice is no longer required. However, in this article, we will still
assume that the twisted charge induced by the D7-branes 
vanishes locally~: in other words, four pairs, each made of D7-branes
with opposite twisted charges, will be located at points  $z=\pm
m_i,\,i=1,2$. Therefore, all the twisted charge will be due to the
presence of fractional D3-branes. 

Separating the D7-branes from the O7-plane generates non trivial
dilaton and untwisted R-R scalar. The presence of these sources for
the R-R field leads to the following asymptotic behaviour~:
\begin{equation}
\tau(z) \sim i \left(1 + \frac{g^B_s}{2\pi} \left(8 \log (z) -
2\sum_{i=1}^2 \log (z^2-m_i^2) \right)   \right)
\label{tauasymp} 
\end{equation}
which displays the expected monodromy. We have also taken
into account a constant background value for $\tau_2$. However,
the function (\ref{tauasymp}) does not always have a positive
imaginary part, and, 
therefore can not be the complete answer. 

This problem has been solved in \cite{gsvy}, where, using its modular 
properties, a formula has been proposed for $\tau$~:
\begin{equation}
j(\tau(z)) = {z^{-8}}{\prod_{i=1}^{2}(z^2-m_i^2)^2}.
\end{equation}
$j$ is the modular invariant function and $\tau$ is restricted
to the fundamental domain ${\cal F}$ of the modular group. Indeed, $j$
induces a one-to-one mapping between ${\cal F}$ and the $z$-plane. At
large $\tau_2$, one obtains the result (\ref{tauasymp}) but, even if
this approximation will be adequate for the purpose of this article,
one should keep in mind that it is not the whole solution and that 
corrections could be needed to tackle other problems. 

The only sources for the twisted field equation (\ref{gammaeq}) are 
the actions (\ref{d3action}) of the two kinds of fractional D3-branes.
The positions of the D-branes of the first kind will be denoted
$\pm z_{1p}$ for $p=1,\ldots,N_1$ while, for the other set, we will use
$\pm z_{2q}$ for $q=1,\ldots,N_2$.  
As in the previous case, this information leads to the formula~: 
\begin{equation}
\gamma(z) \sim 4 i \pi g^B_s \alpha^\prime \,\sum_{\epsilon=\pm
1}\left(\sum_{p=1}^{N_1} \log\frac{(z+\epsilon z_{1p})}{\Lambda} - 
\sum_{q=1}^{N_2} \log\frac{(z+\epsilon z_{2q})}{\Lambda}\right)
\label{gammaasymp} 
\end{equation}
from which we can extract the twisted scalar $b$ and the twisted
4-form $A^{(4)}$ after an Hodge duality~:
\begin{eqnarray}
&&\tilde{b} \sim 4 \pi g^B_s \alpha^\prime \,{\rm Re}\sum_{\epsilon=\pm
1}\left(\sum_{p=1}^{N_1} \log\frac{(z+\epsilon z_{1p})}{\Lambda} - 
\sum_{q=1}^{N_2} \log\frac{(z+\epsilon z_{2q})}{\Lambda}\right), 
 \nonumber\\
&&A^{(4)} \sim 4 \pi g^B_s \alpha^\prime \,{\rm Re}\sum_{\epsilon=\pm
1}\left(\sum_{p=1}^{N_1} \log\frac{(z+\epsilon z_{1p})}{\Lambda} - 
\sum_{q=1}^{N_2} \log\frac{(z+\epsilon z_{2q})}{\Lambda}\right)\, dx^0
\wedge \ldots \wedge 
dx^3.  \nonumber 
\end{eqnarray}
As explained in the appendix of \cite{bdflmr}, this profile only encodes
one-loop open string effects. However, usually \cite{bdflmp,pol}, one
takes 
this formula as the full solution to the field equation. This choice
can be justified if one insists on preserving the rotational symmetry
in the $z$-plane when all the D-branes are located at $z=0$. 
Indeed, other holomorphic solutions involving 
subleading corrections to (\ref{gammaasymp}) breaks this
symmetry, which, from the gauge theory point-of-view, corresponds to
the $U(1)$ of the $R$-symmetry group. As we will see below,
this solution contains enough information about the gauge coupling
renormalization as predicted by open string/closed string duality
and, from the gauge theory point-of-view, we do not expect any higher
order perturbative corrections to 
it. However, this argument does not rule out the existence of
exponentially vanishing corrections of order $(\Lambda/z)^n$, $n>0$,
which could be attributed to D-instanton effects and, under the
gauge/gravity correspondence, are expected to
be mapped Yang-Mills instantons effects. 

Actually, in the context of fractional D2-branes \cite{wz}\footnote{In
the paper \cite{wz}, it was believed that the Harvey-Liu solution
\cite{harveyliu} was 
mapped under heterotic on $T^4$/type IIA on $K3$ duality to the
D6-brane wrapped on $K3$ configuration of \cite{jpp} but, actually, it
corresponds to the fractional D2-brane case. The wrapped D6-brane
configuration is mapped to a superposition of H and KK-monopoles in
the heterotic string and its ``non-abelian'' generalization is not
known so far.}, it has been
argued recently that the leading order solution receives corrections which are
exponentially small at infinity but become important at some finite
distance from the core. 
The idea was to ``improve'' the effective description by going to a
special point of the moduli space of  the type IIA on $T^4/{\mathbf
Z}_2$ string theory  where fractional 
D0-branes become massless. Indeed, the breakdown of the supergravity
description in the case of fractional D2-branes corresponds to the
appearance of new massless states at a special radius $r_e$, due to
fractional D0-branes, which 
had been integrated out in the standard supergravity action. At this
radius, a $U(1)$ gauge theory (corresponding to the R-R 1-form
$A^{(1)}$ under which the 
fractional D2-branes are magnetically charged) gets enhanced to a
$SU(2)$ group, the 
new massless states playing the role of W-bosons. A correct
way to deal with this problem is to go directly to the $SU(2)$ point
of the moduli space of the string theory and search for a solution of
the new 
field equations. In this case, the twisted fields, $b$ and $A^{(1)}$,
obey a $SU(2)$ Bogomolnyi equation. Its analytic solution is known for
a magnetic charge equals to one. Asymptotically, the
solution looks like the Dirac monopole but, close to the
enhan\c{c}on radius, non-perturbative effects due to the fractional
D0-branes completely change the picture and lead to a smooth,
non-singular metric (except obviously at the origin, where we always
have a singularity, as for any D-brane).  
The twisted NS-NS field $b$ plays the role of a Higgs field and, away
from the special radius $r_e$, the $SU(2)$ group is broken to a $U(1)$
by the Higgs mechanism and the fractional D0-branes get a mass
proportional to $b$. 

In the case of fractional D3-branes, we expect that similar effects
will also solve the singularities of the metric, by taking into account
the fact that fractional D-instantons become massless at $r=r_e$. More
precisely, one expects the modified solution for the twisted
fields to be given by the periodic monopole solutions considered by
\cite{ck}, in the limit of a vanishing circle~: in this paper, the
authors discuss solutions to the non-abelian Bogomolnyi equation on
$\mathbb{R}^2 \times S^1$. In the language of the last paragraph, this
would describe fractional D2-branes with a transverse direction
(also transverse to the orbifold)
compactified. In the  zero radius limit, we obtain the T-dual
configuration, namely fractional D3-branes.  Obviously the
asymptotics, given by the 
Dirac solution, coincide with (\ref{gammaasymp}) but, unfortunately, the
full solution is not known, even in the case of a charge one periodic
monopole. However, one expects that the solution will be modified at
short distances, like for the t'Hooft-Polyakov monopole, in such a way
that the metric singularities are smooth out. More speculatively, one
could conjecture that 
the exact solution is such that the cascading-like behaviour of
the D3-brane charge discussed in \cite{pol} and refuted in \cite{bgl},
completely disappears, since we do not expect such a Seiberg duality
in a ${\cal N}=2$ gauge theory. It would be interesting to
investigate these questions further but, for the purpose of this
article, we stop these speculations now and consider only the
formula (\ref{tauasymp})
and (\ref{gammaasymp}). 

Using a D-brane probe, we will show that these
asymptotics carry enough information to reproduce the one-loop corrected
effective action on 
the moduli space of the $Sp(2N_1)$ gauge theory with $2N_2 + 2$
hypermultiplets in the fundamental representation. The Laplace
equation (\ref{heq}) for the harmonic function $H$ can not be solved
explicitly in the case of D7-branes away from the origin of the
$z$-plane but will not be needed.  When the untwisted R-R charge is
zero, one can choose $\psi$=0 and the harmonic function will be
similar to the case of type IIB on $\mathbb{C}^2/\mathbb{Z}_2$
\cite{bdflmp}. 

To probe the background, we consider a fractional D3-brane of the
first kind located at a 
position $z$.  Inserting into its action ${\cal S}_1$ the supergravity
fields, we can extract the gauge coupling and the kinetic term 
for the adjoint scalar fields $\Phi^a$ which belong to the vector
multiplet. The expansion of the action 
to quadratic order in the gauge and scalar fields leads to~:
\begin{eqnarray}
{\cal S} &=& -\frac{1}{8\pi g^B_s} \int\, d^4x e^{-\phi}
\left(1+\frac{\tilde{b}}{2\pi^2\alpha^\prime} \right) \left(
\frac{1}{4} F_{\alpha\beta}F^{\alpha\beta}+
\frac{1}{2}\partial_\alpha{\Phi}^a \partial^\alpha{\Phi}^a \right) 
\nonumber \\ 
&=& -\frac{1}{g^2_{\rm YM}(z)} \int\, d^4x  \left(
\frac{1}{4} F_{\alpha\beta}F^{\alpha\beta}+
\frac{1}{2}\partial_\alpha{\Phi}^a \partial^\alpha{\Phi}^a \right) 
\end{eqnarray}
with 
\begin{eqnarray}
\frac{1}{g^2_{\rm YM}(z)} &=& \frac{1}{8\pi g^B_s} + \frac{1}{8\pi^2} {\rm
Re} 
\sum_{\epsilon=\pm 1}
\left(
\sum_{p=1}^{N_1} 2\log\frac{(z+\epsilon z_{1p})}{\Lambda} - 
\sum_{q=1}^{N_2} 2\log\frac{(z+\epsilon z_{2q})}{\Lambda}\right) \nonumber\\
&&~~~~~~~~~~~~~~~~+ 
4 \log(\frac{z}{\Lambda}) -
\sum_{i=1,2} \log\frac{(z^2-m_i^2)}{\Lambda^2}
\end{eqnarray}
This formula is the one-loop corrected gauge coupling
of the symplectic gauge theory on its Coulomb branch \cite{dkp}, where
the {\it 
vev} of its adjoint scalars are proportional to $z_{1p}$, with $2N_2+2$
fundamental hypermultiplets. 
This one-loop renormalized gauge coupling displays Landau poles where
the enhan\c{c}on mechanism has been argued to take place. However, we
know from the gauge theory point-of-view that the complete moduli
space metric, 
which includes non-perturbative - Yang-Mills instantons - corrections
is a  smooth manifold, something that our solution is not
able to capture.

\subsubsection{$Sp(2N_1)\times SO(2N_2)$ gauge theory}

The set of fixed points of the other orientifold projection, $P_2$,
reduces to the plane left $x^6=x^7=x^8=x^9=0$ left invariant by the
orbifold operator $\alpha$. There is no dilaton tadpole and the
complex scalar $\tau$ is a constant. However, the orientifold
projection induces a background twisted charge \cite{dp,bz,pu}, which is
precisely twice the charge of a fractional D3-brane of the first
kind. One of the consequences of this background charge
is that the conformal field theory case requires fractional D-branes
and not just bulk D-branes~: it is obtained for $N_2 = N_1 + 1$. 
A solution to the twisted field equation which
correctly encodes the presence of these sources reads  
\begin{equation}
\gamma(z) = 4 i \pi g^B_s \alpha^\prime \,\left(\sum_{\epsilon=\pm
1}\left(\sum_{p=1}^{N_1} \log\frac{(z+\epsilon z_{1p})}{\Lambda} - 
\sum_{q=1}^{N_2} \log\frac{(z+\epsilon z_{2q})}{\Lambda}\right) + 2
\log\frac{z}{\Lambda}\right). 
\end{equation}
Again, we can probe this supergravity
background using the two different kinds of fractional D3-branes. 
Extracting the kinetic terms for
scalar fields on the first kind of probe located at $z$ leads to
\begin{eqnarray}
{\cal S}_1 &=& -\left(\frac{1}{8\pi g^B_s}+\frac{1}{8\pi^2} {\rm Re}
\sum_{\epsilon=\pm
1}\left(\sum_{p=1}^{N_1} 2\log\frac{(z+\epsilon z_{1p})}{\Lambda} - 
\sum_{q=1}^{N_2} 2\log\frac{(z+\epsilon z_{2q})}{\Lambda}\right) + 4
\log\frac{z}{\Lambda}\right) \nonumber \\
&&~~~~~~~~~~~~~~~~~~~~~~~~~~~~~~~~~~~~~~~~~~~~\int\, d^4\sigma  \left(
\frac{1}{4} F_{\alpha\beta}F^{\alpha\beta}+
\frac{1}{2}\partial_\alpha{\Phi}^a \partial^\alpha{\Phi}^a \right) 
\end{eqnarray}
while, for the other probe,
\begin{eqnarray}
{\cal S}_2 &=& -\left(\frac{1}{8\pi g^B_s}+\frac{1}{8\pi^2} {\rm Re}
\sum_{\epsilon=\pm
1}\left(\sum_{q=1}^{N_2} 2\log\frac{(z+\epsilon
z_{2q})}{\Lambda}) - 
\sum_{p=1}^{N_1} 2\log\frac{(z+\epsilon z_{1p})}{\Lambda}\right)
 - 4
\log\frac{z}{\Lambda}\right) \nonumber \\
&&~~~~~~~~~~~~~~~~~~~~~~~~~~~~~~~~~~~~~~~~~~~~\int\, d^4\sigma  \left(
\frac{1}{4} F_{\alpha\beta}F^{\alpha\beta}+
\frac{1}{2}\partial_\alpha{\Phi}^a \partial^\alpha{\Phi}^a \right) 
\end{eqnarray}
In these expressions, we recognize respectively the one-loop corrected
metric on the moduli space of the ($Sp(2N_1) + 2 N_2\, \square_1$) and
($SO(2N_2) + 2 N_1\, \square_2$) gauge theories. 

\ 

To conclude this first part, we must mention that there is (at least)
another way 
to obtain fractional D3-branes with orthogonal or symplectic gauge
groups. The construction, which has been described in \cite{uranga} for
the case 
of type IIB string theory on $\mathbb{C}^2/\mathbb{Z}_{2N}$, 
uses an $O3$-plane. For $N=1$,  
the gauge theory which lives on the two possible kinds of fractional
D-branes is $Sp(2N_1)\times SO(2N_2)$ with a hypermultiplet in the
bifundamental. In this paper, one can also see that the orientifold
projection, called $\Omega \Pi$, does not introduce any untwisted
tadpoles but gives the same background twisted charge in the sector
$k=N$ as the $P_2$ projection we considered in this
section. Therefore, the supergravity solution and the gauge couplings
renormalizations  will be identical.

\section{Fractional D-branes versus wrapped M5-branes}

\subsection{D3-branes on a two-centers Taub-NUT space}

Alternative approaches to describe non-conformal pure ${\cal
N}=2$ $SU(N)$ super-Yang-Mills theories using supergravity
solutions have been discussed recently \cite{pw,gkmw,bcz,bfms}. 
It would be interesting to make connections  between these various
models and to completely understand which points on the moduli space
of the gauge theory they describe.  
The aim of this section is to provide a first step in this
direction by comparing the fractional D3-brane supergravity solution
of type IIB on the non-compact orbifold $\mathbb{C}^2/\mathbb{Z}_2$ to
another proposal, namely the M5-brane wrapped on a
Riemann surface of \cite{bfms}. 

The strategy we will adopt is to lift the type IIB solution to
eleven dimensions. To do this, we will first replace the orbifold
singularity by the two-centers Taub-NUT space. Indeed, we know that,
on one hand, a $k$-centers Taub-NUT manifold degenerates into a
$\mathbb{C}^2/\mathbb{Z}_k$ orbifold in a specific limit and, on the other hand,
it is T-dual to $k$ NS5-branes delocalized along the T-dualized
direction \cite{ov,ghm}. Then, we
construct fractional D3-brane solutions on this ALF space, using the
method described  in \cite{bcffst} to obtain the 
supergravity 
solution for D3-branes with fluxes on the Eguchi-Hanson manifold.
Performing
a T-duality along the fibered $U(1)$ of the Taub-NUT manifold, we will obtain a
type IIA solution describing D4-branes stretched between NS5-branes.
Finally, using the type IIA/M-theory duality 
dictionary, we  will lift the solution to eleven dimensions.  

Let us first begin by reviewing some properties of the two-centers
Taub-NUT space \cite{egh}, which will be useful to find the D3-brane
solution.  
The metric of the Taub-NUT manifold parameterized by the coordinates
$(x^6, \ldots, x^9)$ and with centers located at $x^i=x^i_1,\, x^i_2$
for $i=7, 8 ,9$,
can be written as~:
\begin{equation}
ds_{\rm Taub-NUT}^2 = V\delta_{ij} dx^i dx^j + V^{-1} (dx^6 + A_i
dx^i)^2
\label{taubnut}
\end{equation}
with
\begin{eqnarray}
V=1+\sum_{a=1,2}\frac{R}{\vert x^i - x^i_a \vert}  \qquad {\rm and }
\qquad \partial_i V = \epsilon_{ijk} \partial_j A_k. 
\label{vtaubnut}
\end{eqnarray}
In order to avoid conical singularities at $r=0$, the coordinate $x^6$ must
have periodicity $4\pi R$. The metric is asymptotically flat and 
approaches the cylinder $\mathbb{R}^3 \times S^1_{4\pi R}$ at large $|x|$. 
When the centers become coincident, this metric has a $\mathbb{Z}_2$
singularity at $r=0$. In this case, the orbifold limit is obtained
by sending the radius $R$ to infinity. More precisely, if one defines
the new coordinates $(r, \theta, \phi, \psi)$
\begin{eqnarray}
&&x^6 = 2 R \psi,~~~~x^7 = r \cos\theta,~~~~x^8 = r \sin\theta,
\cos\phi~~~~{\rm and}~~~~x^9 = r \sin\theta \sin\phi \nonumber  
\end{eqnarray}
in the limit $R \rightarrow \infty$, the metric (\ref{taubnut}) with
$x_1=x_2=0$ 
degenerates to 
\begin{eqnarray}
ds^2 \simeq 2R \left(\frac{1}{r} 
\left(dr^2 + r^2(d\theta^2 + \sin^2 \theta d\phi^2) \right) 
+ r (d\psi + \cos \theta d\phi)^2 \right)
\end{eqnarray}
where $\psi \sim \psi + 2 \pi$. We recognize the metric of an ALE space
with an $A_1$-type 
singularity.  
If we define the radial coordinate $\tilde{r} = 2\sqrt{2R r}$,
we obtain a flat space metric with a $\mathbb{Z}_2$ identification.  

The two-centers Taub-NUT manifold (\ref{taubnut},\ref{vtaubnut})
possesses a homology 2-cycle 
given by the fibration of the $S^1$ parameterized by $x^6$ over the
segment joining the two centers. For coincident centers, in the limit
$R \rightarrow \infty$, it degenerates to the vanishing 2-cycle of the
$\mathbb{Z}_2$ orbifold
located at the origin of the space. The dual of this cycle is an
anti-self-dual harmonic 2-form, $G^{(2)}$, defined as
\begin{equation}
G^{(2)} =  d(V^{-1}(dx^6+2R\cos \theta d\phi))
\end{equation}
and which satisfies the
relation 
\begin{equation}
G^{(2)}\wedge G^{(2)} = - \Delta(r)\, \Omega_{\rm Taub-NUT}
\end{equation}
where 
$\Delta(r) \equiv (\partial V^{-1})^2$ and $\Omega_{\rm Taub-NUT}$ is
the volume form of Taub-NUT.  

To recover the fractional D3-brane solution in the orbifold limit, we
will search for a D3-brane metric of the form
\begin{equation}
ds_{\rm IIB}^2 = H^{-1/2} dx^2_{1+3} + H^{1/2} \left(dz d\bar{z} +
ds^2_{\rm Taub-NUT} \right).
\end{equation}
induced by the presence of fluxes of NS-NS and R-R 2-forms through
the 2-cycle of the Taub-NUT space\footnote{An
alternative approach would be to search for a ``true'' D5-brane solution
wrapped on the 2-cycle. This has been explored in
\cite{gkmw,bcz}. There, the NS-NS 2-form is zero while the 
dilaton has a non-trivial profile. Moreover, the solution does not
carry any D3-brane charge.}. In the orbifold limit, they become
twisted NS-NS and 
R-R scalars. They are expressed in term of the harmonic 2-form~: 
\begin{equation}
C^{(2)} + i B^{(2)} = \gamma\, G^{(2)}
\end{equation}
and the complex scalar $\gamma$ is a solution of the two dimensional
Laplace equation  
\begin{equation}
\square_{\mathbb{R}^2} \gamma = 0 
\end{equation}
in the space transverse to the D3-branes and to the Taub-NUT
manifold. 
As in the previous section, requiring the preservation of
supersymmetries imposes that $\gamma$ is a holomorphic function. 
Moreover, we would like to have $N$ fractional D3-branes located at
positions $z_p$. A solution satisfying these constraints is given by
\begin{eqnarray}
\gamma(z) = 4 i \pi g^B_s \alpha^\prime \,
\sum_{p=1}^{N} \log\frac{(z+ z_{p})}{\Lambda}
\end{eqnarray} 
but, as discussed in the previous section, it could receive
corrections which improve its behaviour close to the D-branes. 

Finally, the Einstein equation implies that the harmonic function
$H(r,z,\bar{z})$ must verify~:
\begin{equation}
  \left( \square_{\mathbb{R}^2} + \square_{\rm Taub-NUT} \right)  H = - 
 \vert \partial_z \gamma \vert^2 \Delta(r).
\label{einstein}
\end{equation}
A standard way to obtain solutions to this equation is to perform a
Fourier transformation over $z$ and $\bar{z}$. For simplicity, we will
consider only the case of $z_p=0$  for which the rotational symmetry in
the $z$-plane is preserved. We call $\rho =
\sqrt{z\bar{z}}$. Therefore, the Fourier transform of the 
harmonic function $H$ is defined by
\begin{equation}
\hat{H}(r, \mu) = \frac{1}{2\pi} \int_0^\infty d\rho \, \rho \,
J_0(\mu \rho) \, \left(H(r, \rho)-1 \right).
\end{equation}
$\mu$ runs from $0$ to infinity. For convenience, the constant part of
the harmonic function has been subtracted to define its Fourier
transform. 
Then, the field equation (\ref{einstein}) becomes
\begin{equation}
\left(\partial_r^2 + \frac{2}{r} \partial_r - \mu^2
\left(1+\frac{2R}{r}\right) \right) \hat{H} + \hat{J} = 0,
\label{eqfourier}
\end{equation}
where $\hat{J}$ is the Fourier transform of $V(\partial_r V^{-1})^2
|\partial_z \gamma|^2$. 

The general solution of this equation is the sum of the homogeneous
equation solutions and of a particular solution. 
One can see that the homogeneous equation is related to the Whittaker
equation; its generic solution is given by 
\begin{equation}
\hat{H}_h(r, \mu) = \frac{1}{r} \left(\alpha(R, \mu) 
W_{-\mu R,-\frac{1}{2}}(2 \mu r) + 
\beta(R, \mu) W_{\mu R,-\frac{1}{2}}(-2 \mu r)  \right).
\end{equation}
To fix the integration constants $\alpha$ and $\beta$, we will impose
(hopefully) reasonable D3-brane boundary conditions in the
asymptotic regions $r\rightarrow \infty$ ($R$ small) and $r \rightarrow
0$ ($R$ large). 

First, one would like to have an asymptotically flat metric when
$r\rightarrow \infty$. This implies that $\lim_{r\rightarrow \infty}
\hat{H}(r, \mu) = 0$.  On the other hand, we know
the behaviour of the Whittaker function when the norm of its argument
is large \cite{gr}~: 
\begin{equation}
W_{a, b}(x) \sim x^a \, e^{-x/2}~~~~{\rm for}~~~~|x| \rightarrow \infty. 
\end{equation}
This means that $\beta$ must vanish. 

Then, we can consider the ALE space limit, namely $R$ large,
with $\tilde{r}=2\sqrt{2R r}$ fixed, arbitrary, being the radial
coordinate of the four-dimensional flat space. At leading order, we
would like to recover the first term of the harmonic function of
fractional D3-branes \cite{bdflmp} at the orbifold conformal field
theory point 
where they carry half of the mass and of the charge of bulk D3-branes
\footnote{This
expression is the same as for $N$ standard D3-branes in a flat
space. However, one must remember that the $\mathbb{Z}_2$ 
identification $\psi \sim \psi +2\pi$ divides by two the volume of the
surrounding 5-sphere used to compute the (untwisted) D3-brane charge
and, therefore, gives only an $N/2$ D3-brane charge.},  
$$H_{h}\left(\frac{\tilde{r}^2}{{8R}},\rho\right) \simeq 1+\frac{4\pi
N g^B_s 
\alpha^{\prime 2}}{(\tilde{r}^2 + 
\rho^2)^2},$$
whose Fourier transform is expressed in term of the Bessel function
$K_1$ as~: 
$$\hat{H}_{h}\left(\frac{\tilde{r}^2}{8R},\mu\right)
\simeq  \frac{4\pi N g^B_s \alpha^{\prime
2}}{\tilde{r}}\,\frac{\mu K_{1}(\mu\tilde{r})}{2}.$$ Using 
the following property of the Whittaker function \cite{gr},
\begin{equation}
\lim_{R \rightarrow \infty} \left(\Gamma(1+\mu R)\, W_{-\mu R,
-\frac{1}{2}}\left(\frac{\mu \tilde{r}^2}{4R}\right)\right) =
 \mu \tilde{r} K_1(\mu\tilde{r}), 
\end{equation}
one finds that $\alpha(R, \mu) \sim 
4\pi N g^B_s \alpha^{\prime 2}\,{\Gamma(1+\mu R)}/{16  R}$  when $R$ is
large.   A priori, this formula is valid only in the ALE space limit
and does not say anything when $R$ is small.
However, one can try to see if we obtain a sensible answer when we
extrapolate this expression to small $R$, where the Taub-NUT 
space is asymptotically of the form $\mathbb{R}^3 \times S^1_{4\pi R}$. 
 
Therefore, let us consider as the full solution of the homogeneous
equation the following expression~: 
\begin{equation}
\hat{H}_h(r, \mu) = \frac{4\pi N g^B_s \alpha^{\prime 2}\,{\Gamma(1+\mu
R)}}{16Rr}
W_{-\mu R,-\frac{1}{2}}(2 \mu r)
\label{solution}
\end{equation}
and study its small $R$ behaviour. 
If we expand this formula in the asymptotic region $R$ small, 
$r\rightarrow \infty$, we obtain~:
\begin{equation}
\hat{H}_h(r, \mu) \sim \frac{4\pi N g^B_s \alpha^{\prime
2}\,{\Gamma(1+\mu R)}}{16Rr} \, e^{-\mu r}\,(1+\mu R \log(2\mu r)+\ldots) 
\end{equation}
Its leading order in an expansion in series of $R$ can be Fourier
transform back to~: 
\begin{equation}
H_h(r, \rho) \sim 1 + \frac{4\pi N g^B_s \alpha^{\prime 2}}{16 R
(r^2+\rho^2)^{3/2}} 
\label{leadharm}
\end{equation}
In this expression, we recognize the smeared harmonic function
of $N/2$ D3-branes delocalized along the circle $S^1_{4\pi
R}$. We see that the division by two of the D3-brane charge imposed in
the ALE space limit by choosing a fractional D3-brane boundary
condition has propagated to the other limit where the Taub-NUT space
degenerates to $\mathbb{R}^3 \times S^1_{4\pi R}$ with $R$
small. Actually,  when $R$ is sent to zero, the type IIB
description is not the correct one anymore and one must go to its type
IIA dual. In the next section, we will perform explicitly this
T-duality and  we will show that the
division by two confirms the interpretation of the T-dual
configuration as  
D4-branes stretched between NS5-branes, carrying half of the charge
of standard D4-branes.  

However, before doing this, we must mention that the full solution
could be
obtained by adding to (\ref{solution}) a particular solution of the
equation 
(\ref{eqfourier}).  To find such a solution one can use the method of
\cite{bcffst}~: one 
constructs first the Green function $G(\mu| r, r^\prime)$ with the
appropriate boundary conditions using the solutions of the homogeneous
equation. Then, the particular solution will be given by
\begin{equation}
\hat{H}_{n-h}(r, \mu) = - \int_0^\infty G(\mu| r, r^\prime)
\hat{J}(r^\prime, \mu) (r^\prime)^2 dr^\prime. 
\end{equation}

\subsection{Uplift to eleven dimensions and comments}

Using transformation rules for NS-NS and R-R fields given in
\cite{bho}, we can perform a T-duality along the isometric direction $x^6$
to map the above solution to a type IIA one. Discussions in the
literature have argued that the dual of the
two-centers Taub-NUT space is given by two NS5-branes whose
separation in the $x^6$ direction is related to the twisted scalar
$b$ and that the fractional D3-branes become D4-branes stretched
between the two NS5-branes. The T-dual coordinate $\tilde{x}^6$ has
periodicity $\pi\alpha^\prime/R$ and the type IIA gauge coupling constant
is $g^A_s = g^B_s \sqrt{\alpha^\prime} / 2 R$. 
The D4-branes/NS5-branes metric, dilaton, NS-NS and R-R
 field strength read~:
\begin{eqnarray}
&&ds_{\rm IIA}^2 = H^{-1/2} dx^2_{1+3} + H^{1/2} \left(dz d\bar{z} +
V\delta_{ij} dx^i dx^j \right) + H^{-1/2} V (d\tilde{x}^6 - b\, dV^{-1})^2,
\nonumber \\
&&e^{\phi_A} = V H^{-1/2}, \qquad H^{(3)} = \frac{\alpha^\prime}{\tilde{R}}
\sin \theta\, d(\tilde{x}^6 - 
b\, V^{-1})\wedge d\theta \wedge d\phi, \nonumber \\
&& F^{(2)} = - dc \wedge dV^{-1}, \nonumber \\
&& F^{(4)} = \frac{\alpha^\prime}{\tilde{R}} \sin \theta \,  V^{-1}
\left( 
(d\tilde{x}^6 - b \, dV^{-1}) \wedge dc  + 2 
c\, db \wedge dV^{-1}
\right)\wedge d\theta \wedge d\phi, \nonumber \\
&& F^{(6)} = dH^{-1} \wedge dx^0 \wedge \ldots \wedge dx^3 \wedge
(d\tilde{x}^6 - b \, dV^{-1}), \nonumber \\
&&\gamma(z) = (c + i b)(z) = 4\pi i N  g^A_s
\frac{(\alpha^\prime)^{3/2}}{\tilde{R}} \log 
\frac{z}{\Lambda},
\label{tdsol}
\end{eqnarray}
where we have defined the dual radius $\tilde{R} =
\alpha^\prime/2R$. As always after T-duality, the $p$-forms appear at
the same time as the dual $10-p$-forms. However, one can
transform the 4-form $F^{(4)}$ to its Hodge dual and combine the
result with the $F^{(6)}$ of (\ref{tdsol}). More surprising is the
existence of a 
non trivial R-R 2-form $F^{(2)}$. It comes from the dualization of the
2-form potential $C^{(2)}_{x^6r}$ which was required to obtain a
solution which preserves supersymmetry. Dualizing this 2-form
field strength to an 8-form gives~:
\begin{equation}
F^{(8)} = (HV)^{-1} dx^0 \wedge \ldots \wedge dx^3 \wedge db \wedge
H^{(3)}. 
\end{equation} 
Therefore, we see that the R-R 7-form potential vanishes and the
8-form is induced by the non-trivial NS-NS 
3-form field strength $H^{(3)}$ and R-R 5-form potential
$C^{(5)}_{x^0x^1x^2x^3\rho}$. 

Since the type IIA solution has been obtained using T-duality in the
$x^6$ direction, the two NS5-branes are delocalized along this coordinate
and their harmonic function is smeared~: $$V=1+\frac{\alpha^\prime}{\tilde{R}
r}.$$ In particular, momentum modes along $\tilde{x}^6$ have been
neglected \cite{ghm}. Moreover, this NS5-brane harmonic function does
not include the back-reaction due to the stretched D4-branes. 

Now, we can express the leading order
(\ref{leadharm}) of the harmonic function in term of type IIA
variables~:
\begin{equation}
H_h(r, \rho) \sim 1+\frac{\pi N g_s^A (\alpha^\prime)^{3/2}}{2
(r^2+\rho^2)^{3/2}}.  
\label{leadiia}
\end{equation}
This harmonic function corresponds to $N$ D4-branes with half of the
mass and charge of standard D4-branes as one can expect for D4-branes
stretched 
between NS5-branes located at antipodal points of the
circle\footnote{The approximation (\ref{leadiia}) corresponds to
this symmetric configuration since the bending of the NS5-branes due to the
stretched D4-branes appears only when one takes into account the
non-homogeneous solution.}. This result is an argument in favor of the
conjecture that the expression (\ref{solution}) is correct not only
for large $R$ but also for any radius.  

Then, we can uplift this solution to eleven dimensions. The type IIA
and eleven dimensional metrics are related by
\begin{equation}
ds^2_{11d} = e^{-2\phi_A/3} ds^2_{\rm IIA} + e^{4\phi_A/3} (dx^{10} +
C^{(1)}_\mu dx^{\mu})^2
\end{equation}
which leads to the result
\begin{eqnarray}
ds^2_{11d} &=& (HV)^{-1/3} dx^2_{1+3} + (H V)^{2/3} \left(dr^2 + r^2
d\Omega^2_2 \right) 
+ H^{2/3} V^{-1/3} dz d\bar{z} \nonumber \\ 
&&~~~~~~~~~~~~~~+ H^{-1/3} V^{2/3}\left((d\tilde{x}^6-b\,
dV^{-1})^2+(dx^{10}+c\, 
dV^{-1})^2\right).
\label{11dmetric}
\end{eqnarray}
To relate this solution to the wrapped M5-branes configurations
considered in \cite{bfms}, we rewrite the metric as a
function of the new complex coordinate $z^2=s$ defined as~:
\begin{equation}
s \equiv \tilde{x}^6 + i x^{10} + i \gamma(z)\,V^{-1}(r). 
\end{equation}
Renaming $z$ into $z^1$, the metric takes the form of the
Fayyazuddin/Smith ansatz \cite{fs1,fs2}~:
\begin{eqnarray}
ds^2_{11d} &=& g^{-1/3} dx^2_{1+3} + g^{2/3} \left(dr^2 + r^2
d\Omega^2_2 \right) + g^{-1/3} g_{m\bar{n}}dz^m dz^{\bar{n}}
\end{eqnarray}
where $g = H \, V$ is the  determinant of the K\"ahler metric
\begin{eqnarray} 
g_{z\bar{z}} = H + V^{-1} \vert\partial \gamma\vert^2, \qquad
g_{s\bar{s}} = V,\qquad 
g_{z\bar{s}} = -i \partial \gamma, \qquad g_{s\bar{z}} = i
\bar{\partial} \bar{\gamma}.
\end{eqnarray}
It is easy to check that this metric verifies the equations~: 
\begin{equation}
\partial_m \partial_{\bar{n}} g + \square_{3} g_{m\bar{n}} = 0. 
\label{kahler}
\end{equation}
In particular, the equation (\ref{kahler}) for $(m, \bar{n})=(1,
\bar{1})$ is just a 
rewriting of (\ref{einstein}).

To summarize, starting with the type IIB supergravity solution which
describes 
fractional D3-branes on a two-centers Taub-NUT space, we
have obtained a partially smeared supergravity solution representing
M5-branes wrapped on a two dimensional surface parameterized by the
coordinates $s$ and $z$. Since the 
gauge theory living on the world-volume of the D3-branes is a pure
${\cal N}=2$ $SU(N)$ SYM theory, the eleven dimensional solution
should also correspond to this theory and  one expects that the
the Riemann surface is given by the Seiberg-Witten curve of this gauge
theory.  

This problem has been studied directly in eleven dimensions
in \cite{bfms}, where a completely localized solution has been proposed. 
Let us remind in few words their result. The authors of \cite{bfms}
considered a
configuration of M5-branes wrapped on the Riemann surface defined by 
\begin{equation}
f(t,z) = t + 2 B(z) + \frac{1}{t} = 0 
\label{seibergwitten}
\end{equation}
which is the Seiberg-Witten curve of the pure ${\cal N}=2$ $SU(N)$ SYM
theory when $B(z)=\prod_{p=1}^N (z-z_p)$ \cite{wit}. In
this case, only the eleventh coordinate 
is compactified $x^{10} \sim x^{10} + 2\pi R^\prime$ and  $t=e^{
s/R^\prime}$. They consider a 
``decoupling'' limit,  where field theory quantities are held fixed
while the 
eleven dimensional Planck scale is sent to zero. The new variables
$Z$, $S$ and $u$ defined by 
\begin{eqnarray}
Z = \frac{z}{\alpha^\prime} = \frac{z R^\prime}{l^3_P}, \qquad S =
\frac{s}{R^\prime}, \qquad u^2 = \frac{r}{l^3_P}
\end{eqnarray}
are fixed under the scaling. 
In term of these new variables, the K\"ahler potential which defines
the metric 
appears as the sum of two terms
$K=K^{(1)}(u,F(Z,S),\bar{F}(\bar{Z},\bar{S})) + |G(Z,S)|^2$~:
\begin{eqnarray}
&&K^{(1)}(u,F(Z,S),\bar{F}(\bar{Z},\bar{S})) = \frac{\pi N}{u^2} \ln
\frac{\sqrt{u^4+|F|^4}+u^2}{\sqrt{u^4+|F|^4}-u^2}, \qquad
F=f^{1/N},\nonumber \\ 
&&G(Z,S) = - \frac{Z N f^{1-\frac{1}{N}}}{4}  \int_0^1
\frac{dt}{\sqrt{\left(\frac{f}{2} - B(t Z) \right)^2-1}}.
\end{eqnarray}
Using this K\"ahler potential, one can calculate the determinant of
the metric~:
\begin{eqnarray}
g=\frac{\pi N}{2\left(u^4+|F|^4\right)^{3/2}}. 
\label{bfmssol}
\end{eqnarray} 
This metric is singular at the location of the M5-brane, namely on the
Seiberg-Witten curve at $r=0$. However, it does not have the repulsive
singularities displayed by other ${\cal N}=2$ solutions. It would be
interesting to probe this background to see if one can reproduce the
running of the coupling constant on the world-volume of the probe. To
do this, one should consider the action of M5-brane wrapped on the
Riemann surface defined by (\ref{seibergwitten}). 

Unfortunately, we see that the two solutions are obtained in two very
different approximations. Indeed, in the D3-brane case, $s$ parameterizes a 
2-torus while the Seiberg-Witten curve (\ref{seibergwitten}) is valid
only when $\tilde{x}^6$ is a non-compact coordinate. 
Moreover, the M5-brane solution 
considered here is smeared over this 2-torus, whereas the solution
(\ref{bfmssol})  of
\cite{bfms} was completely localized. Since the functions $V$
and $\gamma$ are smeared and depend explicitly on the radius of
compactification, the decompactification limit is meaningless.  The
last difference is that our 
metric is asymptotically flat while  the M5-brane of \cite{bfms} is
only a ``near-horizon'' solution. One can take also a near 
horizon limit in (\ref{11dmetric}) but the other approximations and
hypothesis make the comparison impossible. Obviously, it would be very
interesting to find a solution which goes beyond the limits of  
these two solutions, namely which is fully localized on the 2-torus
and is not just valid in a decoupling region  but the task seems to be 
difficult.   
A better understanding of the properties of the solution of
\cite{bfms}, in particular in the optic of the gauge/gravity
correspondence, is also desirable.

\

\noindent{\bf Acknowledgements} 

I would like to thank M. Bianchi and A. Sagnotti for having given me
the opportunity to present these results to the String Theory group in
{\it Tor Vergata} and for 
discussions. I also wish to thank C. Bachas for the invitation to
visit the Laboratoire de Physique Th\'eorique de l'\'Ecole Normale
Sup\'erieure and for his remarks on this work.  
This research is supported by a PPARC fellowship.

\


\end{document}